\documentclass[twocolumn,secnumarabic,amssymb, nobibnotes, aps, prl, superscriptaddress]{revtex4-1}

\usepackage{float}
\usepackage{graphicx}
\usepackage{subfigure}
\usepackage{amsmath} 
\usepackage{epstopdf}
\usepackage{setspace}
\usepackage[shortlabels]{enumitem}
\usepackage{mathrsfs}
\usepackage{subfig}
\usepackage{color,soul}
\usepackage[dvipsnames]{xcolor}
\usepackage{dcolumn,amssymb,subfig}
\usepackage{bm}
\usepackage{ulem}

\renewcommand{\vec}[1]{\bm{#1}}
\newcommand{\tens}[1]{\mathbf{#1}}

\newcommand{\fig}[1]{\textbf{Fig.\ref{#1}}}

\begin{document}
	\title{Role of friction in multidefect ordering}
	\author{Kristian Thijssen}
	\author{Mehrana R. Nejad}
	\author{Julia M. Yeomans}
	\affiliation{The Rudolf Peierls Centre for Theoretical Physics, Department of Physics, University of Oxford, Parks Road, Oxford OX1 3PU, UK}
	
\begin{abstract}
We use continuum simulations to study the impact of friction on the ordering of defects in an active nematic. 
Even in a frictionless system, +1/2 defects tend to align side-by-side and orient antiparallel reflecting their propensity to form, and circulate with, flow vortices. 
Increasing friction enhances the effectiveness of the defect-defect interactions, and defects form dynamically evolving, large scale, positionally and  orientationally-ordered structures which can be explained as a competition between hexagonal packing, preferred by the -1/2 defects, and rectangular packing preferred by the +1/2 defects. 
\end{abstract}
	
	\maketitle
	

Active materials are out-of-equilibrium systems that continuously consume energy and exert stress on their environment \cite{RevModPhys.85.1143}.
Examples include bacterial suspensions \cite{Volfson08,DoostmohammadiPRL2016,Li19}, living cells \cite{Duclos2017,Saw2017,Kawaguchi2017} and vibrating granular rods \cite{galanis2010}.
The continuous injection of energy - or activity -  and the resulting stress can lead to phenomena such as collective motion~\cite{Dombrowski2004,Sanchez2012,Sumino2012}, and active turbulence \cite{Wensink2012,Dunkel2013},
behaviours which cannot be captured by conventional equilibrium statistical mechanics
\cite{narayan2007long,solon2015pressure,grafke2017spatiotemporal,liebchen2015clustering,ramaswamy2010mechanics}.

\begin{figure*}[t] 
\centering
\includegraphics[width=0.95\textwidth]{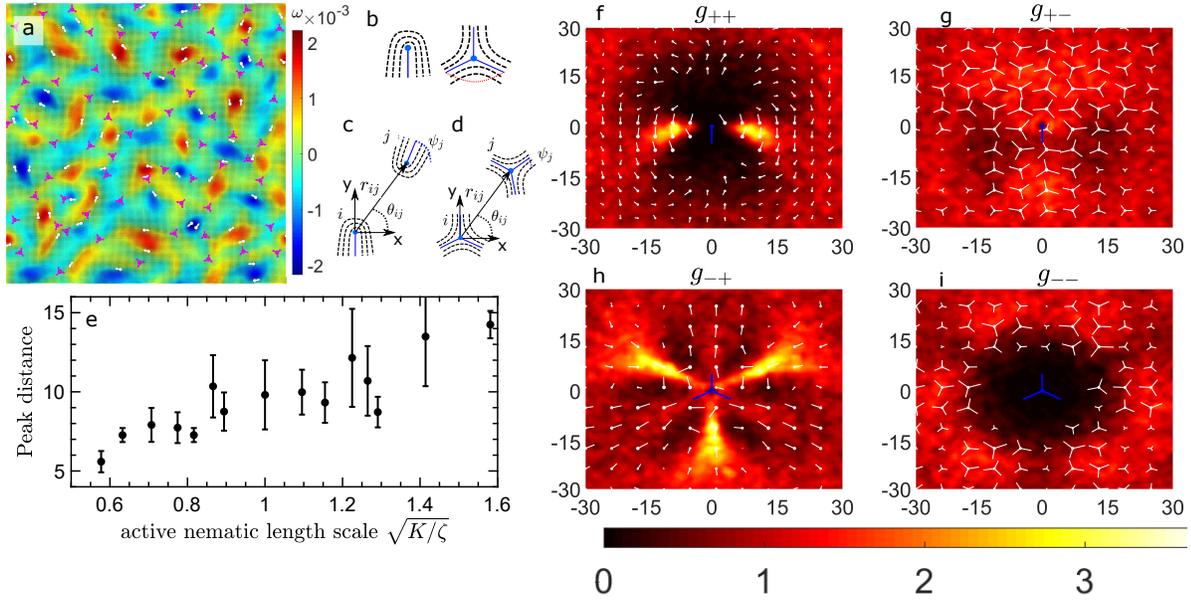}
\caption{	
\textbf{Defect ordering in wet active turbulence:}
(a) Snapshot of active turbulence for very low friction, $F=0.023$. The white (magenta) symbols are $+1/2$ $(-1/2)$ defects. Background colour denotes the vorticity.
(b) Schematic representation of $+1/2$ and $-1/2$ defects. $+1/2$ defects have a single polar axis (blue line) and $-1/2$ defects have three axes.
(c,d) For a reference $+$ (c) or $-$ (d) defect $i$ we define an associated polar co-ordinate system. 
(e) Spacing between $+1/2$ defects (defined as the position of the maximum in $g_{++}$ (in lattice units) as a function of the active length scale $\sqrt{K/\zeta}$. Activity $\zeta$ and elasticity $K$ were varied.
(f-i) Pair distribution function $g_{ab}(r,\theta)$ where $a$ and $b$ represent $+$ and/or $-$ defects showing the positional distribution of $b$-type defects around an $a$-type defect. The white arrows represent the orientational distribution vector $\vec{S}$ with arrow size normalized by the magnitude of $\vec{S}$ and axes are in lattice units.
}
\label{fig:wet_system}
\end{figure*}

Many active systems have nematic symmetry, and such active materials 
extend the physics of passive nematics \cite{doostmohammadi2018active}.
The activity destroys long-range nematic order, resulting in the proliferation of topological defects in the orientation field. 
Moreover, in active systems gradients in the director field induce stresses and hence $+1/2$ topological defects are self-propelled \cite{giomi2013defect,ourprl2013,giomi2014defect}.
Flows driven by the defects, and by other gradients give rise to active turbulence (\fig{fig:wet_system}a), a chaotic flow state characterised by short-range nematic order, high vorticity and localised bursts of velocity 
\cite{thampi2014vorticity,urzay2017multi}.

A key experimental system for investigating the properties of active turbulence is a dense suspension of microtubules propelled by two-headed kinesin motors \cite{Sanchez2012,Guillamat16}.
Investigation of the defect motion in a thin layer of this material showed that the $+1/2$ topological defects can themselves display long-range nematic order while retaining their motile nature \cite{decamp2015orientational}. Very recent simulations of active nematics with hydrodynamics, \textit{wet systems}, have shown short-range defect ordering \cite{kumar2018tunable,pearce2020scalefree}.
Simulations and analytical approaches to active nematics with strong friction, have ignored viscous stress and reproduced $+1/2$ defect ordering, but this is polar rather than nematic \cite{putzig2016instabilities,srivastava2016negative,shankar2019hydrodynamics}.
Such polar defect ordering has been attributed to arch-like configurations of the nematic director field \cite{patelli2019understanding}. In another study in the same regime, rotational contributions of the flow are ignored, and a static lattice of defects with positional and orientational order has been observed \cite{oza2016antipolar}. A lattice was also observed in the high friction regime in the presence of viscous stress \cite{doostmohammadi2016stabilization}.

To clarify how defects order in wet active nematics, we perform large scale continuum simulations to measure both the positional and the orientational order of topological defects with varying friction. 
We confirm that $+1/2$ defects  prefer to position themselves side by side and align anti-parallel \cite{kumar2018tunable, pearce2020scalefree},  while the $-1/2$ defects prefer to impose a three-fold symmetry on their surroundings. 
Increasing friction decreases the hydrodynamic screening length, which measures the {competition} between viscocity and friction, and increases the effectiveness of the defect-defect interactions, and  the defects start to form dynamically evolving orientationally and positionally ordered structures 
even in the regime where defects are still motile. This
can be explained in term of the competition between hexagonal packing preferred by the $-1/2$ defects and rectangular packing preferred by the $+1/2$ defects. 
The range of the ordering increases with increasing friction 
in agreement with experiments \cite{decamp2015orientational}.




To investigate the orientational arrangements of defects, we solve the continuum equations of motion for a 2D active nematic~\cite{Saw2017,thampi2014vorticity} using a hybrid Lattice Boltzmann method \cite{marenduzzo2007steady,B908659P,lbdriven,hemingway2016correlation,kruger2017lattice,Guillamateaao1470,doostmohammadi2017onset,carenza2019lattice,hardouin2019reconfigurable,PhysRevLett.124.187801}.
This is now well documented, so we summarise relevant points here, giving the full equations and simulation details in the Supplemental
Materials \cite{suppmat}. 
The relevant hydrodynamic variables are an orientational order parameter $\tens{Q}$, which describes the magnitude and direction of the nematic order, and the velocity.
 We consider low Reynolds number  
and work above the nematic transition temperature, so any nematic order is induced solely by the activity, and consider a flow-aligning fluid. 
The equations of motion are identical to those describing the nematohydrodynamics \cite{DeGennes,beris1994thermodynamics} of passive nematic liquid crystals except for an additional term in the stress $-\zeta \tens{Q}$ which implies that any gradients in the nematic ordering drive flows and, for extensile activity, $\zeta>0$, results in active turbulence \cite{Sriram2002}. 
Lastly, we include a friction coefficient $f$ in the Navier-Stokes equation modelling energy loss from the 2D active plane to its surroundings.

\noindent\textit{Defect distributions:-}         
    To measure positional and orientational correlations between defects, we treat the $+1/2$ and $-1/2$ defects as two different types of quasi-particle with different symmetries (\fig{fig:wet_system}b)~\cite{vromans2016orientational}. 
Defects are found by measuring the local winding number \cite{hobdell1997numerical,vcopar2013visualisation} (see SM for details \cite{suppmat}).
We consider a reference defect $i$ and choose a co-ordinate system with the reference defect as the origin and the Cartesian axes oriented relative to the defect as shown in \fig{fig:wet_system}c,d.  To define the relative position of the second defect, we use polar co-ordinates $(r,\theta)$ defining $\theta$ as the angle from the $x$-axis. We measure the relative position of the other defects $j$ present at a given time step (\fig{fig:wet_system}c,d), and then sum over all the measured defect pairs, taking data every $1,000$ time steps to get the pair-wise positional distribution function: 
    \begin{equation}
g_{\pm\pm}(r,\theta)=\frac{V}{N_{\pm\pm}}\sum_t  \sum_{\pm \pm \mbox{pairs}} \delta(r-r_{ij},\theta-\theta_{ij}),
\label{eq:eq1}
    \end{equation}
	where the subscripts of $g$ indicate the type of the defect pair $ij$, e.g.~${-+}$ refers to the positioning of $+1/2$ defects around a $-1/2$ defect. The normalization ${V}/{N_{\pm\pm}}$ is the area divided by the total number of defect pairs $N_{\pm\pm}$. 
 We introduce this normalization to set $g=1$ at $r\rightarrow \infty$ to normalize to bulk densities at large distances. To acquire sufficient statistics each distribution function is based on measurements of at least $10^6$ defect pairs which requires runs $\sim 3$ orders of magnitude longer than the average defect lifetime (see SM \cite{suppmat}).
	
In addition to the relative defect positions, we are also interested in the average defect orientation relative to the reference defect. To obtain this information, we calculate the orientation distribution vector,
\begin{eqnarray}
&&\vec{S}_{\pm\pm}(r,\theta)= \nonumber \\
&&\mathcal{N_{\pm\pm}}\sum_t  \sum_{\pm \pm \mbox{pairs}}    \delta(r-r_{ij},\theta-\theta_{ij})
\begin{bmatrix}
\cos{\kappa_j\psi_j}\\ 
\sin{\kappa_j\psi_j}
\end{bmatrix}
\!\!,
\label{eq:eq2}
\end{eqnarray}
where $\psi_{j}$ is the polar angle of the orientation of defect $j$ in the co-ordinate frame defined by the reference defect $i$  (\fig{fig:wet_system}c). Here $\kappa_j=2(1-k_j)$, where $k_j$ is the charge of the $j$th defect, accounts for the three-fold rotational symmetry of the $-1/2$ defects. Taking the normalization constant as $\mathcal{N_{\pm\pm}}=V/(N_{\pm\pm}g_{\pm\pm}(r,\theta))$ means that the magnitude of $\vec{S}$ is $0$ in the absence of orientational correlations and $1$ if the defect orientations are perfectly correlated. To avoid statistically insignificant data, we set $g=0$ and $\vec{S}=0$  if the defect count for any site  is lower then 5.



	\begin{figure}[t] 
		\centering
		\includegraphics[width=0.45\textwidth]{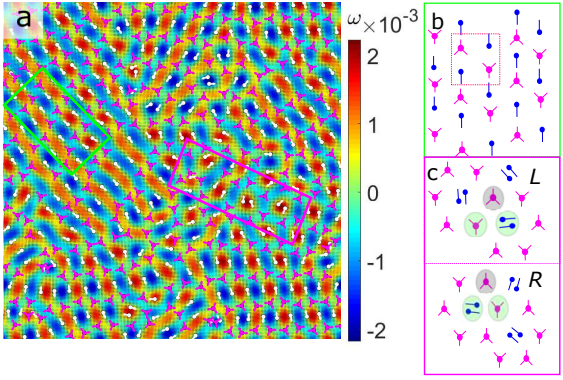}
		\caption{(a) Snapshot of the defect structures at intermediate friction $F=0.023$. $+1/2$ ($-1/2$) defects are shown in white (magenta).  There is transient local defect ordering into a rectangular (green outline) or a hexagonal (magenta outline) pattern. The background colour represents the vorticity field.
		(b) Schematic of the rectangular ordering.
		(c) Schematic of the hexagonal ordering. This is chiral: $-1/2$ defects (in grey) have either a left or right neighbouring $-1/2$ defect (in green).  The other position is filled by rotating $+1/2$ defects resulting in local zero charge.
		}
		\label{fig:friction}
	\end{figure}

	\begin{figure}[t] 
		\centering
		\includegraphics[width=0.5\textwidth]{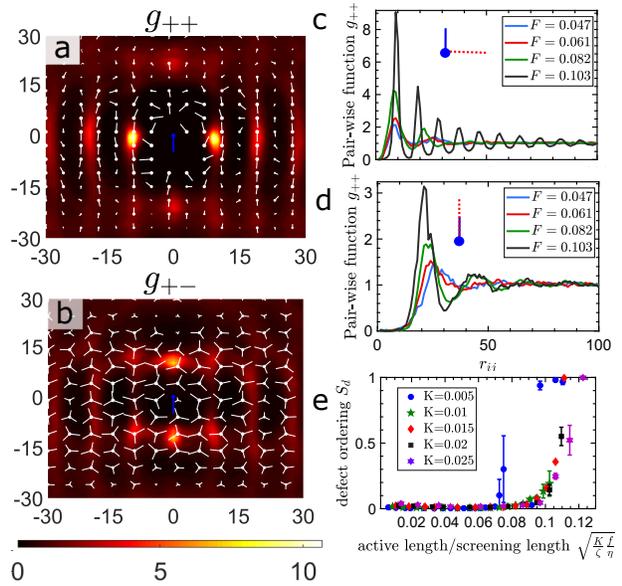}
		\caption{\textbf{Ordering of the $+1/2$ defects at high friction}
		(a-b) Pair distribution function $g_{++}(r,\theta)$ and $g_{+-}(r,\theta)$ (colormap) and the orientation distribution vector (white arrows) for high friction $F= 0.103$. Axes are in lattice units.
		(c-d) $g_{++}(r,0)$ and $g_{++}(r,\pi/2)$ showing the build-up of order along the $x$- and $y$-axes (in the direction of the red dotted line) with increasing friction.
		(e) Nematic defect ordering $S_d$, as a function of dimensionless friction $F=\sqrt{\frac{K}{\zeta}\frac{f}{\eta}}$, for varying elastic constant $K$.
		}
		\label{fig:positive}
	\end{figure}

	\begin{figure}[t] 
		\centering
		\includegraphics[width=0.5\textwidth]{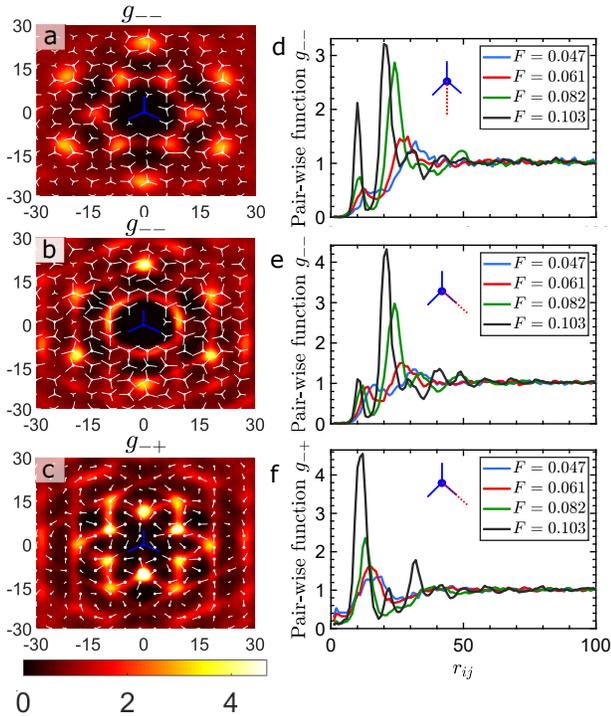}
		\caption{\textbf{Ordering around a $-1/2$ defect at high friction} 
			(a) Pair distribution function $g_{--}(r,\theta)$ (colormap) and the orientation distribution vector (white arrows) for intermediate friction $F= 0.083$.  Axes are  in lattice units.
			(b-c) Pair distribution functions $g_{--}(r,\theta)$ and  $g_{-+}(r,\theta)$ for high friction $F= 0.103$.
			(d-f) $g_{--}(r,-\pi/2)$, $g_{--}(r,-\pi/6)$ and $g_{-+}(r,-\pi/6)$  (along the red dotted line) with increasing friction. 
		}
		\label{fig:negative}
	\end{figure}

\noindent\textit{Emergent defect ordering at low friction:-} 	
We first consider very low friction and high activity, recovering well-developed wet active turbulence (\fig{fig:wet_system}a). 
\fig{fig:wet_system}f shows how positive defects behave in the vicinity of another positive defect: even in this highly turbulent regime there is a clear preference for neighbouring $+1/2$ defects to line up along the $x$-axis in an anti-parallel configuration with a preferred distance between neighbours.
This preferred  defect spacing  scales with the active nematic length scale, $\sqrt{{K}/{\zeta}}$ (\fig{fig:wet_system}e). 
Therefore we choose to measure the friction in terms of a dimensionless friction number $F=\sqrt{(K/\zeta)(f/\eta)}$ which is the ratio of the active length scale to the hydrodynamic screening length.


\fig{fig:wet_system}i shows that $-1/2$ defects prefer not to lie too close to each other, and that there is no preferred length scale in contrast to the $+1/2$ defects.
Interestingly, the $-1/2$ defects do impose an orientational structure on surrounding $-1/2$ defects even in this fully active turbulent regime. 
We already find six peaks where the neighbouring defects have a strong preference for anti-parallel alignment. This is due to the elastic torque \cite{tang2017orientation}. However, the symmetry of the peak positions is caused by the flow fields, which form six vortices around negative defects.
Finally, \fig{fig:wet_system}g,h show that positive and negative defects are preferentially found close together and aligned in the relative orientation associated with creation and annihilation events.


\noindent\textit{Defect lattices at high friction:-} 	
As the friction is increased to $F\sim 0.08$ 
the defect interactions result in large-scale ordering of the defects. As an example, \fig{fig:friction}a presents a snapshot of the defect structure and corresponding vorticity field for $F= 0.083$, where the mean speed of the flow has been reduced by an order of magnitude with respect to the no friction regime.
This figure and movie 1 show that $+1/2$ defects have a strong tendency to form anti-parallel pairs, which induce and orbit on vortices, as already apparent in the no friction limit.
But much larger-scale defect arrangements
also become apparent 
at high friction, as not only the interactions between  the $+1/2$ defects but also those between the $-1/2$ defects result in significant ordering. 
To investigate this, we first consider the structure formed by the $+1/2$ defects (\fig{fig:friction}b), and then the ordering preferred by the $-1/2$ defects (\fig{fig:friction}c).

\fig{fig:positive}a,b show distribution functions of $\pm1/2$ defects around a $+1/2$ defect at strong friction ($F= 0.103$). The first obvious feature of these correlations is that the anti-parallel ordering of the $+1/2$ defects along $x$ is  more pronounced and longer ranged than in the frictionless limit. This is confirmed by \fig{fig:positive}c where we plot the pair-wise  positional distribution function $g_{++}(r,0)$ showing how the strength and range of the correlations increase with increasing friction.

 \fig{fig:positive}d shows that $+1/2$ defects are also ordered along the $y$-axis. This ordering can be interpreted by  
comparing the distribution functions in \fig{fig:positive}a,b which show that  $+1/2$ and $-1/2$ defects alternate along the $y$-axis, and that they align parallel. The ordering increases with friction, but is less pronounced than that along $x$.
We show in the SM that the energy of two $+/-$ defect pairs each arranged as in \fig{fig:friction}b and held at a fixed distance apart, is minimised if the pairs line up along the $y$-axis \cite{suppmat}. Moreover, this configuration is favoured because it leads to non-conflicted flows. We note that this result relies on the presence of intervening -1/2 defects, and is different from the active torque between two $+1/2$ defects studied in \cite{shankar2018defect}.
Together, the preferred ordering of $+1/2$ defects along $x$ and $y$, i.e. perpendicular and parallel to the polar axis of the defects, is satisfied by the rectangular packing of defects shown in \fig{fig:friction}b.


In \fig{fig:positive}e we plot the nematic order parameter, $S_d=-1+2 \sum_t \sum_{++pairs}(\hat{m}_i \cdot \hat{m}_j)^2/N_{++}$ where $\hat{m}_j$ is the polar axis of the $j$th $+1/2$ defect, as the friction and elastic constants are varied. The data collapse confirms $F$ as a suitable control parameter for the simulations. We find that $S_d$ takes a non-zero value, even when the defects are still motile, and increases with increasing friction. It is reminiscent of the experimental system of microtubules driven by motor proteins where the nematic order of defects increases with decreasing film thickness \cite{decamp2015orientational}. However, the patterning also exhibits higher-order symmetry than just nematic as the ordering of defects is polar or anti-polar depending on their relative positions. Upon increasing the friction further ($F=0.106$ in \fig{fig:positive}e), the defects stop moving and a vortex lattice with orientational defect order is established \cite{doostmohammadi2016stabilization} 
on scales comparable to the system size, which is $\sim 15$ times the active length scale. To check whether this is a true transition, we ran simulations on larger lattices which showed that the ordering decreases with increasing system size (reported in the SM \cite{suppmat}). Thus, at these values of $F$, we observe coexisting domains with long- but not infinite-range order. At yet higher frictions the dynamics becomes too slow to {allow feasible simulations of the defect lattices and, for $F \gtrsim 0.14 $, the activity is too weak to create defects.}



\fig{fig:negative} presents results for the ordering around negative defects showing a distinct difference between intermediate ($F=0.082$), \fig{fig:negative}a,d) and high friction  ($F= 0.103$) \fig{fig:negative}b,e). In the intermediate friction regime there are six first neighbour and six second neighbour peaks in the  positional distribution function around the central defect, corresponding to a hexagonal packing of $-1/2$ defects. Both right-handed and left-handed lattices are possible (see \fig{fig:friction}c and Movie 1).
With increasing friction, however, the nearest neighbour peaks become less pronounced showing that it is increasingly difficult to form a hexagonal lattice.

Instead the secondary peaks become more pronounced. The reason for this is  apparent from \fig{fig:negative}c,f, which shows that the $+1/2$ defects increasingly line up along the polar arms of the $-1/2$ defects, and lie between two $-1/2$ defects 
\cite{shankar2018defect,pearce2020scalefree}. We show in the SM that this is the elastically preferred configuration of two defect pairs \cite{suppmat}.  It corresponds to the polar ordering of alternate +1/2 and -1/2 defects seen in the rectangular lattice (\fig{fig:friction}b and Movie 1). 

\noindent\textit{Conclusion}:-
We have numerically investigated defect ordering in an active nematic with hydrodynamic interactions and increasing
 friction. We show that friction can introduce nematic ordering of defects on length scales many times larger then the active length scale, as observed in experimental systems \cite{decamp2015orientational}. A local measurement would, however, give polar order of $+1/2$ defects in the direction of their polar axis (mediated by intervening -1/2 defects), and anti-polar order of the $+1/2$ defects perpendicular to this axis. 

Weak signatures of this ordering are observed even in fully developed active turbulence with no friction. Upon increasing the friction they result in structures with longer-ranged order.  The $-1/2$ defects tend to reorganize themselves into hexagons, where each hexagon encompasses two $+1/2$ defects which rotate on a vortex.   However, this  is not an ideal configuration for pairs of $\pm 1/2$ defects and, as a consequence, the hexagonal packing of defects coexists with
the rectangular structure shown in  \fig{fig:friction}b. 
As the friction is increased, and the hydrodynamic screening length becomes comparable to the active length scale, the rectangular packing becomes dominant, and the system eventually freezes into the rectangular lattice   \cite{doostmohammadi2016stabilization,oza2016antipolar} . 
	
	\section*{Acknowledgements}
	We thank Amin Doostmohammadi for fruitful discussions. K.T. received funding from the European Union's Horizon 2020 research and innovation programme under Lubiss the Marie Sklodowska-Curie Grant Agreement No. 722497. M. R. N. acknowledges the support of the Clarendon Fund Scholarships. 
	
	\bibliographystyle{apsrev4-1}
	\bibliography{references}

\end{document}